# Characterization and suppression techniques for degree of radiation damping in inversion recovery measurements

Andrew Keeler[1,*], Heather M. Whitney[1]

[1]Department of Physics, Wheaton College, Wheaton, IL 60187, USA

**Abstract**: Radiation damping (RD) has been shown to affect T1 measurement in inversion recovery experiments. In this work, we demonstrate that the extent of RD depends upon the T1 of the sample. RD difference spectroscopy (RADDSY) is used to characterize the severity of RD, while gradient inversion recovery (GIR) is used for RD suppression in T1 measurements. At 9.4 T, for the radiation damping characteristic time ($T_{rd}$) of 50 ms, these investigations show non-negligible RD effects for T1 values greater than $T_{rd}$, with severe distortions for T1 longer than about 150 ms, showing reasonable agreement with the predicted $T_{rd}$. We also report a discrepancy between published expressions for the characteristic RD time.

## I.   Introduction

The longitudinal relaxation rate, T1, is a useful quantitative measurement of the state of protons in a material, including as a biomarker for biophysical status. It can also be utilized in other experiments that use the exponential recovery governed by T1 to estimate parameters of interest such as those associated with magnetization transfer. However, measuring it accurately should involve anticipating the presence of radiation damping (RD) and adapting experimental methods that can ameliorate the effects of it (Eykyn et al. 2005).

---

[*] Current address: Department of Physics, University of Tennessee, Knoxville, TN, USA.

RD is a phenomenon that results from the electromagnetic coupling between the transverse magnetization and the current induced in the receiver coil. It was first described in the middle of the twentieth century (Bloembergen & Pound 1954), but was only seen as an irrelevant curiosity until much more recently with the advent of much higher-field and higher-sensitivity NMR systems. The current in the receiver coil, via Lenz's law, produces an additional magnetic field in the sample that is proportional to the transverse magnetization and is 90° out of phase with regard to the transverse magnetization for a well-tuned probe. This field, called the RD field ($B_{rd}$), produces a torque on the magnetic moment on the sample that will rotate the magnetic moment back towards the equilibrium position, providing an additional pathway for realignment with $B_0$ other than relaxation. The RD pathway is characterized by a time constant $T_{rd}$, which is dependent on the experimental setup and given by Eq. 1,

$$T_{rd} = (2\pi\gamma f Q M_0)^{-1} \qquad \textbf{Eq. 1}$$

where $Q$ is the quality factor of the probe and $f$ is the filling factor, the fraction of the coil that is filled by the sample (Abragam 1961; Szantay & Demeter 1999).

In magnetic resonance experiments, longitudinal magnetization re-grows according to an exponential recovery governed by T1. The equilibrium magnetization is considered to have completely re-grown after a time of 5×T1 has elapsed. In order to measure T1, inversion recovery (IR) can be used. The IR technique involves applying an 180° pulse to a sample, completely inverting the equilibrium magnetization. The magnetization is then allowed to relax back towards equilibrium for a particular delay time, after which a 90° pulse is applied to detect by how much the magnetization has relaxed back to equilibrium during that delay. By varying the delay length, a recovery curve for the magnetization can be plotted which follows Eq. 2, given by

$$M_z(t) = M_0 \left(1 - 2 * e^{-\frac{t}{T1}}\right). \quad \text{Eq. 2}$$

An inversion pulse can be made selective by appropriate choice of length of time of inversion (Edzes & Samulski 1978; Gochberg & Gore 2003). Typically, either the integral or peak of the NMR data can be used to fit measurements to this model, although this is complicated by the presence of RD, as discussed later.

When $T_{rd}$ is much longer than T1 the RD effects are negligible and can be safely ignored. However, when $T_{rd}$ is much shorter than T1, the RD processes dominate and the recovery deviates markedly from the expected exponential recovery, exhibiting an abrupt, broken recovery curve (Eykyn et al. 2005). Other well-known effects of strong RD include broadening of the line width for simple spectrum NMR experiments (affecting the assessment of T2*), narrowed linewidths for pulse flip angles approaching 180 degrees and symmetrical negative side lobes for NMR spectra (Mao & Ye 1993), as well as so-called 'fish-shaped' FID signals that will attain a maximum at some point after a pulse is applied but before decaying completely (Szantay & Demeter 1999). It should be noted that it is possible to amplify and make use of the effects of RD for the purposes of solvent suppression, but this is outside of the scope of this work (Augustine 2002). Our approach seeks to suppress RD in experiments in order to obtain more accurate quantitative relaxation measurements. We employ this investigation specifically to draw attention to the fact that severity of RD depends upon the T1 of the sample under investigation. In this work, we characterize the extent of RD of samples of a range of T1 values and demonstrate the use of a pulse sequence to address it.

The first step in utilizing a RD suppression scheme for magnetic resonance experiments is to characterize the severity of RD in the experiments in question. If RD produces only negligible effects on the recovery scheme, then it becomes much less important to find a suppression technique and also much more difficult to show whether any particular suppression technique is effective. To characterize the range of possibilities over which RD has an effect, it becomes much more useful to compare T1 with $T_{rd}$. Szantay et al. (Szantay & Demeter 1999) use T1 in conjunction with $T_{rd}$ and T2* to delineate different regimes of RD effects, from no effects where $T_2^* < T1 < T_{rd}$, to significant effects where $T_{rd} \ll T_2^* < T1$. Of particular note in this characterization scheme are the regimes of weak ($T_2^* < T_{rd} < T1$) and mild ($T_{rd} < T_2^* < T1$) RD. These areas show the effects of RD, but the effects are very slight and can be easily missed. The broken nature of the inversion recovery, for example, only becomes apparent for some mild cases of RD, and the linewidth difference of weakly-damped spectra might only experience broadening of a fraction of a Hz, making such lineshape diagnostic aids only useful in cases of more extreme cases of RD. In this work, because of the effects of radiation damping on the assessment of T2*, we aim to categorize the extent of RD without the use of T2*, focusing on the calculation of $T_{rd}$ and measurement of T1.

Calculating $T_{rd}$ from Eq. 1 is a fairly straightforward process, since the only factor that varies by sample is $M_0$, which can be calculated via a determination of the proton density and use of the relationship given in Eq. 3, as

$$M_0 = \frac{\rho \gamma^2 \hbar^2 B_0}{4 k_b T} \qquad \text{Eq. 3}$$

where $\rho$ is the proton density of the sample, $\gamma$ is the gyromagnetic ratio, $\hbar$ is the reduced Planck's constant, $B_0$ is the magnetic field strength, $k_B$ is the Boltzmann constant, and $T$ is the absolute temperature.

There is a difficulty here, though. The formula for $T_{rd}$ is cited in two forms in the literature. The most frequently cited form is Eq. 1. However, Mao (Mao & Ye 1997) cites the relationship given in Eq. 4, which is

$$T_{rd} = \frac{2 \varepsilon_0 c^2}{\gamma f Q M_0} \qquad \text{Eq. 4}$$

where $c$ is the speed of light in a vacuum and $\varepsilon_0$ is the permittivity of free space. Mao clearly notes that this form makes use of SI units. However, applying SI units to Eq. 1 with the exception of $\gamma$, which is referred to in MHz/T in Abragam shows that Eq. 4 is larger than Eq. 1 by a factor of ten (Appendix A), and seems to generate a $T_{rd}$ value that is inconsistent with our observed RD artifacts as described below, while Eq. 1 gives a value of $T_{rd}$ that is more consistent with the observed RD artifacts.

However, Eq. 1 makes use of non-SI units, making it impossible to combine simply with Eq. 3. For this reason, it becomes important to propose a modification of Eq. 4 as the better expression for determining $T_{rd}$ for a given experimental setup. The modified expression is given as

$$T_{rd} = (5\mu_0 \gamma f Q M_0)^{-1} \qquad \text{Eq. 5}$$

where the extra factor of 10 has been removed and the substitution $\epsilon_0 c^2 = \frac{1}{\mu_0}$ has been made, with $\mu_0$ being the permeability of free space. This formulation is equivalent to Eq. 1, but has the additional advantage of utilizing SI units, making it easier to utilize alongside other mathematical formulae.

However, since there are other factors that might mimic the effects of some RD artifacts, such as so-called 'transition-band effects' generated by an over-filled NMR tube (Szantay & Demeter 1999), an empirical test was needed to estimate the value of $T_{rd}$ and determine which form of the equation was the correct one. The empirical test chosen is referred to as Radiation Damping Difference Spectroscopy (RADDSY) (Szantay & Demeter 1999), and aims to detect RD by measuring the effect of $M_{xy}$ on the recovery of $M_z$. This is done by comparing the $M_z$ recovery curve generated following a 90° pulse with the $M_z$ curve generated via a saturation recovery pulse sequence. The pulse sequences for these two recoveries are as follows:

(1)          $90° - \tau - G - 90° - FID$

(2)          $90° - G - \tau - 90° - FID$

(1) is the pulse sequence for the recovery affected by RD, (2) is the saturation recovery pulse sequence, $\tau$ is a time delay that is varied to generate the recovery curve, and $G$ is a z-gradient that is applied to destroy any transverse magnetization. $G$ immediately following the initial 90° pulse will remove the $M_{xy}$ from the sample, leaving nothing for RD effects to act on, so this recovery will be unaffected by RD. Having a delay before applying $G$ leaves $M_{xy}$ to generate RD effects, while $G$ is used later to eliminate any possible distortions from having both longitudinal and transverse components of magnetization following the 90° 'read' pulse.

In samples where any sort of RD effects are present, there will be a more rapid initial recovery of $M_z$ than there would be in un-damped experiments. This can be seen by plotting the difference between the damped recovery and the un-damped recovery. For strong damping levels, the difference rises sharply with the damped recovery and then falls away as the un-damped recovery catches up and reaches equilibrium. For insignificant levels of RD, the difference curve remains nearer to zero during the course of the recovery, though there is still a slight initial boost in the damped recovery. RADDSY measurements are especially helpful for the middle category, when comparison of T1 and $T_{rd}$ without the use of T2* is needed.

Having noted both empirical and theoretical methods by which the severity of RD in NMR experiments may be assessed, the next step is to establish a method by which RD may be suppressed in experiments where those distortions are present. Suppression techniques in recovery experiments may be broadly divided into two different classes of techniques. The first are termed hardware solutions in which the current induced in the receiver coil is suppressed by the NMR system directly. This can be accomplished by active electronic feedback (Broekaert & Jeener 1995) or utilizing a Q-switch (Anklin et al. 1995). The second class of suppression techniques modifies existing pulse sequences to use rf pulses that destroy any transverse magnetization prior to acquisition. We will utilize gradient inversion recovery (GIR), which uses a magnetic field gradient to continuously suppress any transverse magnetization during the evolution time (Sklenar 1995). There is a third class of suppression techniques, but these techniques are used to suppress RD during spectroscopy by greatly attenuating the transverse magnetization during acquisition (Michal 2010; Khitrin & Jerschow 2012) and are not useful for relaxometry experiments.

Hardware solutions may be practical for some facilities, but many locations such as primarily undergraduate institutions lack the technical resources to modify the NMR spectrometer in the ways necessary to enact such suppression techniques. As such, pulse sequence modifications seem the most practical to investigate in light of their relative ease of application and similar levels of success in prior literature.

## II. Materials and Methods

### A. Sample preparation

All magnetic resonance experiments were carried out at 300 K on a Bruker Ascend 400 MHz spectrometer in 5 mm NMR tubes of samples of purified water doped with CuSO$_4$•5H$_2$O. In order to generate a wide range of T1 values, differing amounts of CuSO$_4$•5H$_2$O were mixed into different samples, which decreases the relaxation times of the samples (Morgan & Nolle 1959; Viola et al. 2000). A range of concentrations approximately 10 to 100 mM was used. In all experiments, a delay time was used that allowed for full recovery of the magnetization, based upon five times the T1 of the sample.

As noted above, for paramagnetic samples, which are often used to create materials of known longitudinal relaxation, the proton density and thus the bulk magnetization of each sample can be calculated using the concentration of the solute. The proton density of pure water is $\frac{2*N_A}{0.000018\ m^3} = 6.7*10^{28}\frac{H}{m^3}$ where $N_A$ is Avogadro's number, but the proton density of CuSO$_4$ solutions is not so readily calculated since the presence of the CuSO$_4$ ions will slightly affect the amount of water present, while Cu$^{2+}$ and So$_4^{2-}$ ions contribute nothing to the proton density. While a fairly reasonable first-order approximation would assume that the proton density for the samples remains roughly constant, a slightly better approximation can be had by assuming that the solute does not change volume in the mixing of each solution. In other words, each solution contains an amount of water equal to the volume of the dry solute subtracted from the total solution volume of 100 ml.

Now consider a 1 L volume of a solution with concentration [n]. In that volume, there will be *n* moles of solute, which will displace *n\*m/d* liters of solvent, where *m* is the molar mass of the solute and *d* is the density in g/L. Also, let $\rho_{h2o}$ be the proton density of pure water and $\rho_{solute}$ be the proton density of the solute. This means that the relationship

$$\rho_{solution} = \rho_{solute} * \left(\frac{[n] * m}{d}\right) + \rho_{h2o} * (1 - \frac{[n] * m}{d}) \qquad \text{Eq. 6}$$

will give the total proton density of the 1 L solution. Of note here is that there is no volume dependence in the calculation of the proton density, so Eq. 6 can be applied to an arbitrary volume of solution.

Combining Eq. 3 and Eq. 6 gives **Eq. 7**,

$$M_0 = \frac{\gamma^2 \hbar^2 B_0}{4 k_b T} [(\rho_{solute} - \rho_{solvent}) \frac{[n] * m}{d} + \rho_{solvent}] \qquad \text{Eq. 7}$$

which can be combined with Eq. 5 to give an expression for $T_{rd}$ in terms of the solution concentrations. Note that **Eq. 7** has been generalized to give the proton density of any single-solute solution, under the assumption that the solution volume is equal to the sum of the dry solute volume and the solvent volume.

Combining Eq. 5 and **Eq. 7** gives Eq. 8,

$$T_{rd} = \left(\frac{4k_b T}{5\mu_0 \gamma^3 \hbar^2 B_0 f Q}\right)\left(\frac{d}{[n]M(\rho_{solute} - \rho_{solvent}) + d\rho_{solvent}}\right) \quad \text{Eq. 8}$$

relating $T_{rd}$ of the solution directly to the concentrations of the solutions and the proton densities of the constituent substances.

## B. Characterization of Radiation Damping

The RADDSY pulse sequence, as described above, was implemented on all samples. A 5% z-gradient was applied over the course of 2 ms. The NMR spectra were integrated due to the effects of RD on lineshape and the difference between the damped and undamped data was evaluated to assess extent of radiation damping.

## C. Suppression of Radiation Damping:

The GIR pulse sequence, as described above, was implemented on all samples. A 5% z-gradient was applied over the entire evolution period. This is important during the GIR experiment since the inverted magnetization in an IR experiment can evolve into a transverse magnetization, giving rise to the RD phenomena that are supposed to be suppressed by the gradient (Mao & Ye 1997). By applying the gradient continuously rather than in a short pulse, any evolution of the magnetization into the transverse plane will be continuously de-phased, keeping $M_{xy}$, and by extension $B_{rd}$, at a minimum. The integral of the resulting NMR spectrum was used for regression to the model given in Eq. 2. Non-linear regression analysis was performed with the use of the MATLAB function nlinfit.

The outputs of the nlinfit function were used in MATLAB's nlparci function to generate a 95% confidence interval for each of the parameters used in the fitting function. For the T1 values obtained this way, the 95% confidence interval corresponds to an error bar with a width of 3.92 standard deviations centered on the T1 value used by the best-fit function.

### III. Results

### A. Characterization of radiation damping

The $T_{rd}$ value for each sample was calculated using Eq. 8. For Q of an estimated value of 100, $T_{rd}$ is approximately 50 ms for all concentrations. The $T_{rd}$ values are equal down to the order of $10^{-4}$ ms, far more precision than is required to define the relaxation regime in which RD has significant effects on recovery. The RADDSY experiments, particularly the difference in the damped and undamped recoveries, were used to qualitatively assess the extent of RD apart from the determination of T2*. Results of the RADDSY experiments are shown in Figure 1.

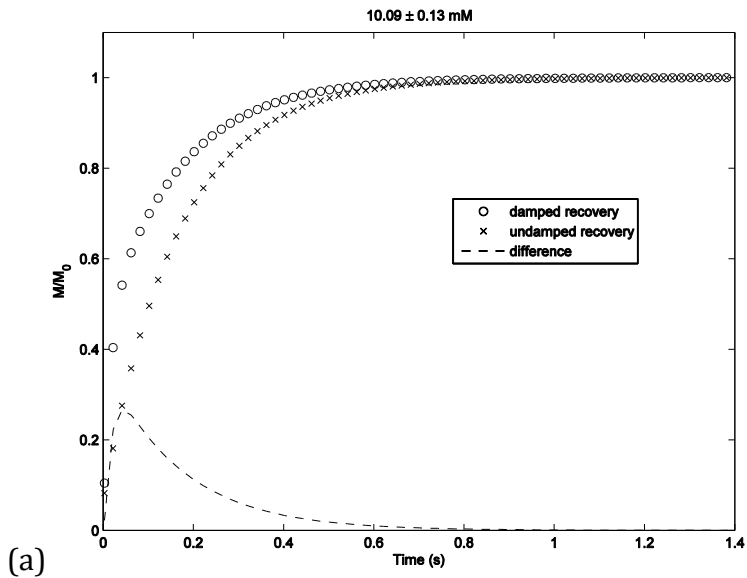

(a)

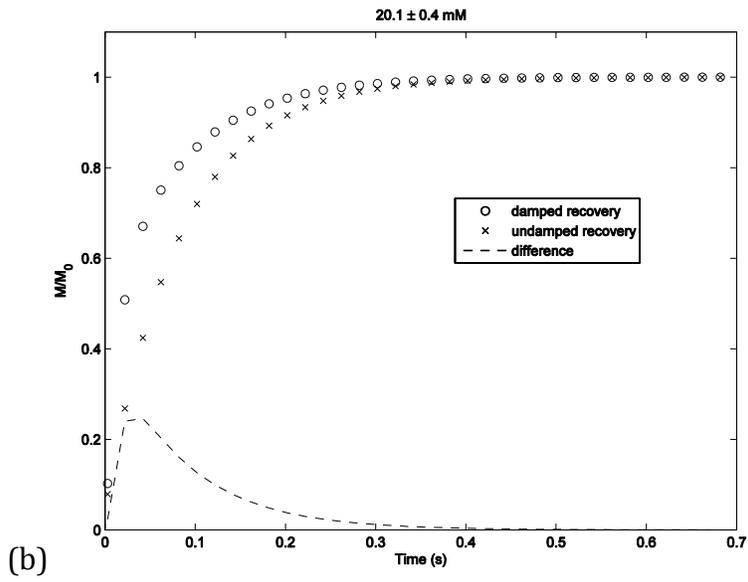

(b)

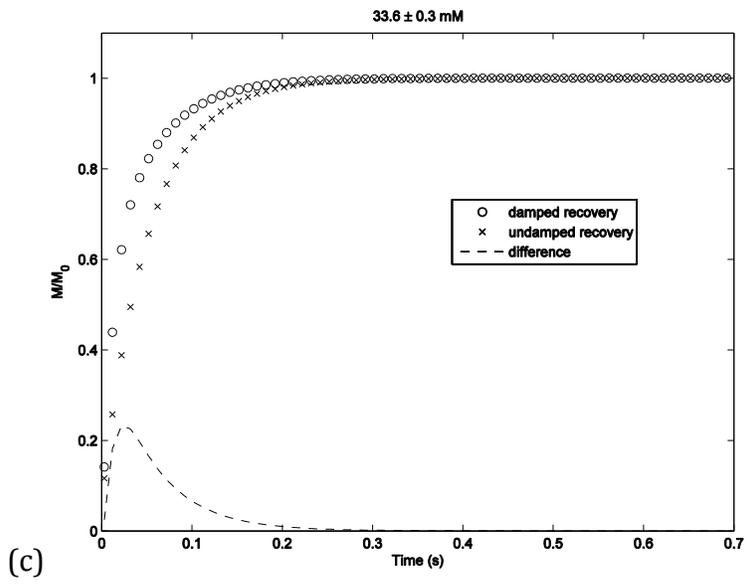

(c)

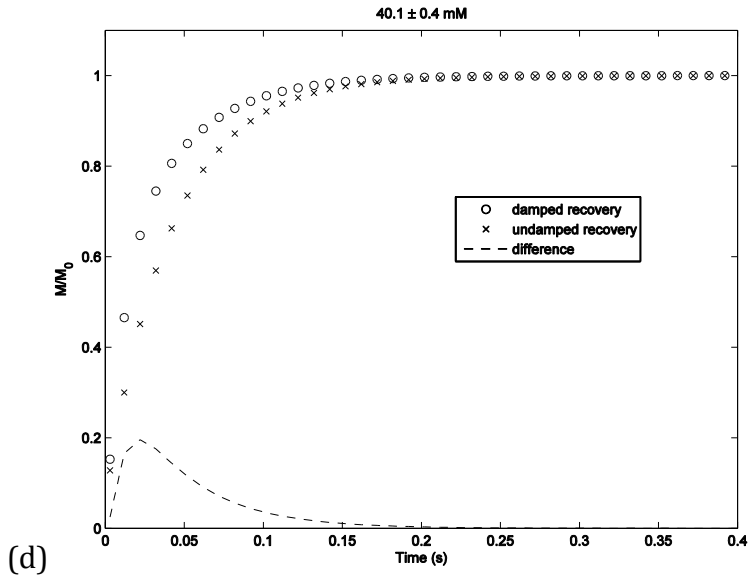

(d)

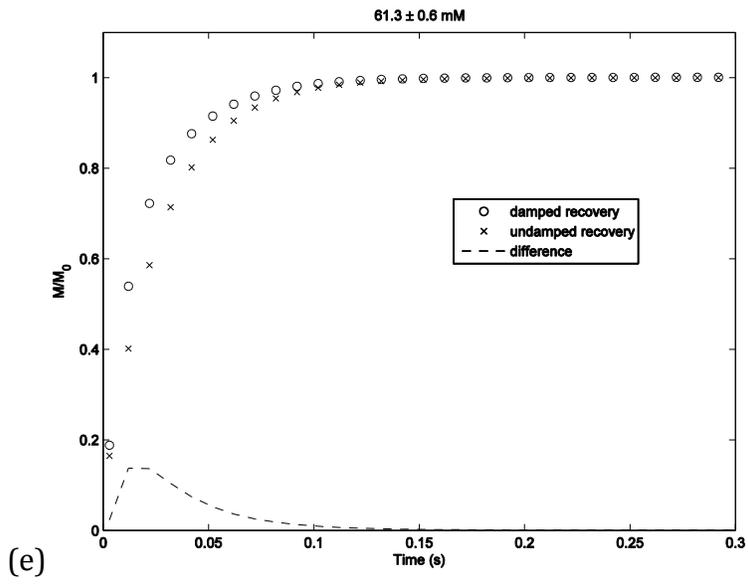

(e)

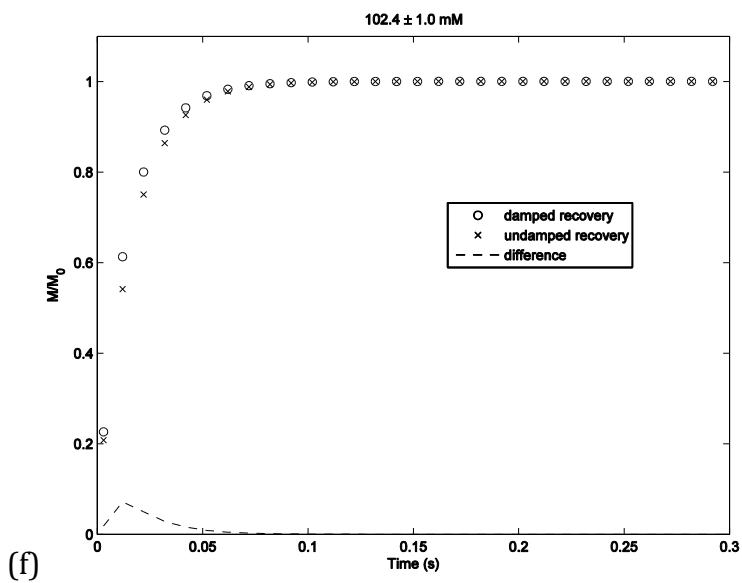

(f)

**Figure 1: RD difference spectroscopy test results for the range of concentrations. The extent of RD can be demonstrated by the decreasing maximum of the difference as concentration increases (and T1 decreases). Of note is the appreciable presence of RD in (c), where the value for T1 (given in Table 1 below) does not suggest a mild or weak level of RD, even though the effects are obviously present in the RADDSY results.**

While the RADDSY experiment has not yet been correlated with degree of RD, these results demonstrate that, particularly in the range of T1 values that are less than $T_{rd}$ but not definitely much less than the value, some RD can be seen.

## B. Suppression of radiation damping

The results of the GIR experiments on selected samples are shown in Figure 2. We focus on concentrations associated with definitely strong and weak RD (Figure 2a and 2c, respectively) as assessed by the Szantay scheme and a concentration for which the RD assessment is less clear using the Szantay scheme (Figure 2b).

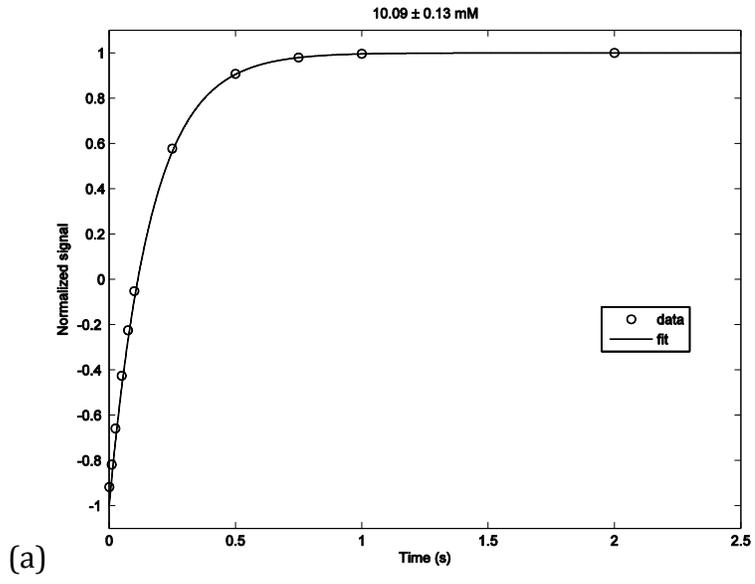

(a)

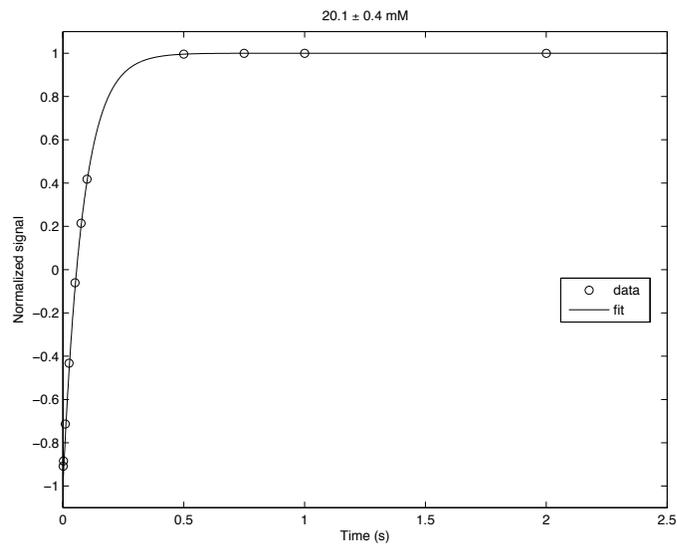

(b)

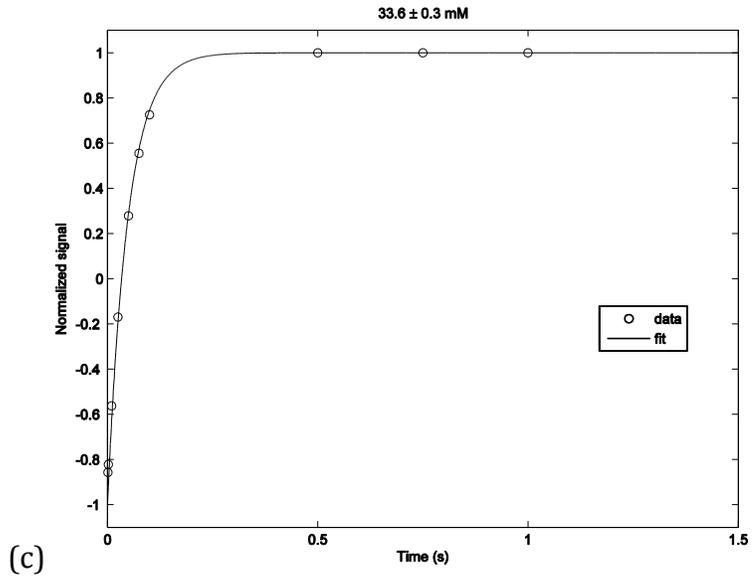

(c)

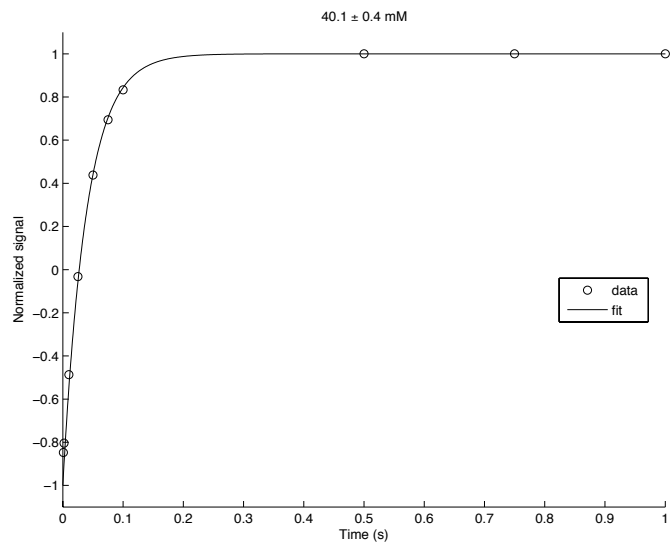

(d)

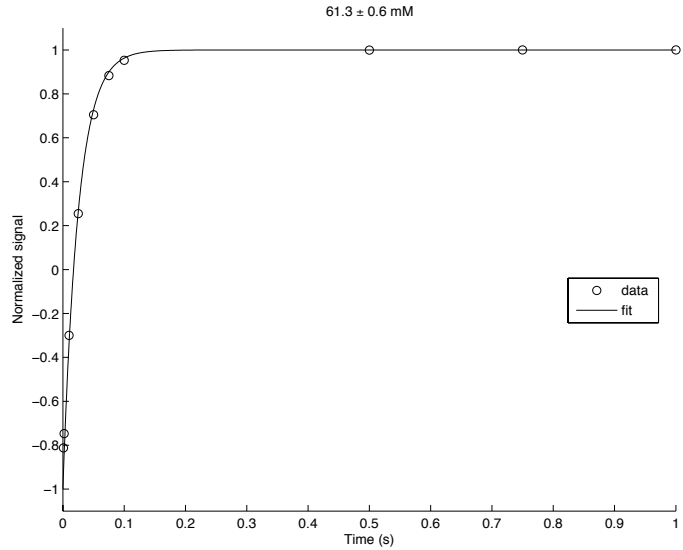

(e)

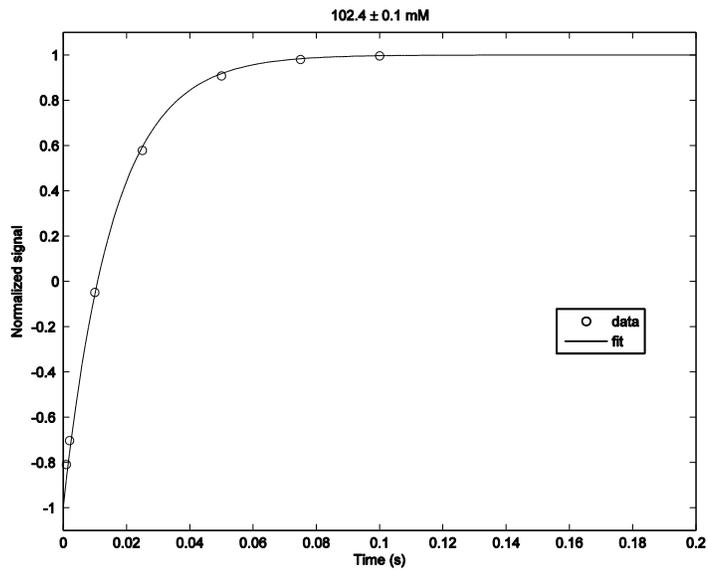

(f)

**Figure 2: Inversion recovery data gathered from GIR experiments on samples of a range of RD. The data is fitted to the model in Eq. 2. The GIR experiment has ameliorated RD from all three regimes, demonstrating that it can be used on samples that include a wide range of T1 values.**

The GIR results allow for the measurement of T1 without the interference of the $T_{rd}$ recovery, and the concentrations of the samples, along with the measured T1 value from the GIR experiments and a characterization of extent of RD, are given in Table 1.

**Table 1: Molar concentrations of doped water samples, T1 measurement, and degree of RD. The concentration error is based on errors of 0.5 mL for solution volume and 0.002 g for solute masses, except for the 20 mM sample, which had a mass uncertainty of 0.004 g and the 100 mM sample, which had a mass uncertainty of 0.02 g. The integral of the NMR spectrum was used for each data point in the regression process for a single-parameter exponential recovery.**

| $CuSO_4$ Concentration (mM) | T1 (ms) | Characterization of RD |
|---|---|---|
| 10.09±0.13 | 154 ± 14 | mild to strong |
| 20.1±0.4 | 79 ± 7 | weak to mild |
| 33.6±0.3 | 48 ± 6 | weak |
| 40.1±0.4 | 38 ± 5 | absent |
| 61.3±0.6 | 25 ± 4 | absent |
| 102.4±1.0 | 15 ± 2 | absent |

As an additional assessment of the accuracy of the measurements, we plotted our measured values of R1, defined as $(T1)^{-1}$, against their respective concentrations of $CuSO_4$ solution. Bucciolini et al. (Bucciolini et al. 1986) predicts that $R_1$ will vary linearly with concentration, while we would expect that RD would increase the R1 values for low concentrations, leading the data points to depart from the linear model near the origin. Figure 3 displays this data, showing that the linear variation prediction is upheld and that there is no significant departure from this plot that matches the predicted effects of RD.

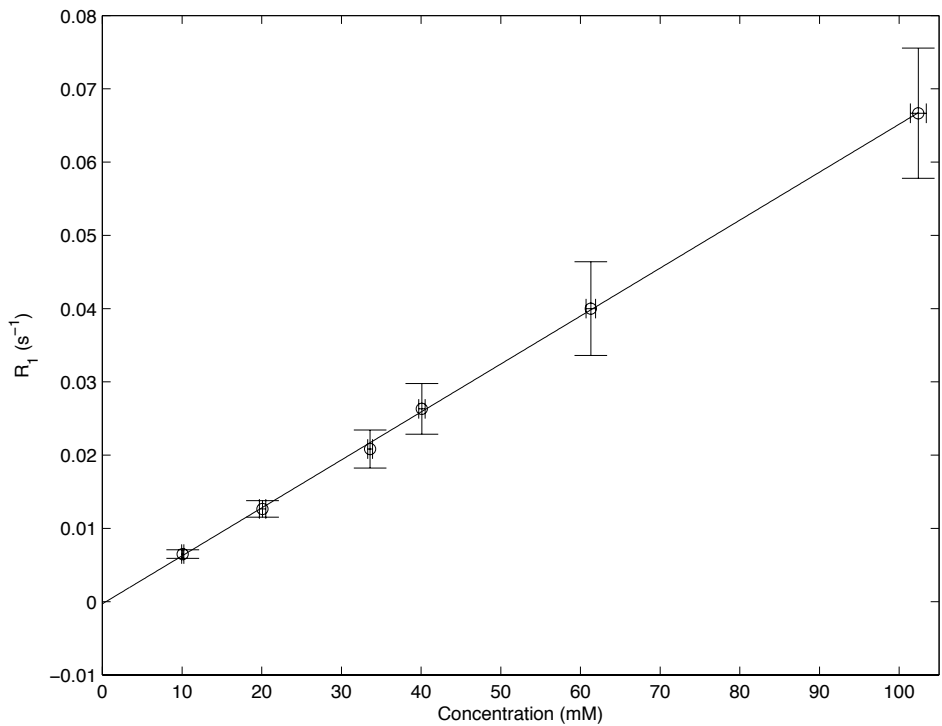

**Figure 3: R1, measured from regression analysis after GIR experiment, versus concentration. The data shows good agreement with a linear model.**

## IV. Conclusions

The results of this investigation show that RD can be a significant factor affecting magnetic resonance inversion recovery experiments, and it should be noted that the severity of RD varies with T1. For NMR systems similar to the one on which this experiment was conducted, at 400 MHz, and the concentrations of doped water, a $T_{rd}$ value of approximately 50 ms is expected, with the primary variations in the $T_{rd}$ value of the system coming from the quality factor of the probe and the proton density of the sample under consideration. A modification to the expression for calculating $T_{rd}$ is proposed (Eq. 8) that reconciles a minor difference in published expressions for the value while incorporating variables for solute concentration.

The RADDSY test is successfully employed to assist in qualifying the effects of RD on recovery experiments, particularly in regimes where the comparison of T1 and $T_{rd}$ is difficult, and the GIR pulse sequence is demonstrated to be effective in suppressing the effects of RD in T1 measurements for a range of T1 values. It is our hope that this work offers detailed information regarding the methods needed to account for RD for quantitative NMR experiments.

There are several extensions of this work that could prove valuable. This investigation deals with the effects of RD for a range of T1 values, but does not look into the effects of RD in other NMR experiments, such as experiments that measure magnetization transfer. Additionally, extensions of the RADDSY test could assist in further being able to definitively assess the extent of RD without explicit measurements of T2*.

**Acknowledgements**

The authors are grateful to the Wheaton College Summer Research Program for its support of this collaborative research.

**Appendix A**: Demonstration that Eq. 1 and Eq. 4 are not equivalent.

Start with Eq. 4. Replace $\varepsilon_0 c^2$ with $\mu_0^{-1}$, since $\frac{1}{\epsilon_0 \mu_0} = c^2$, which gives

$$T_{rd} = \frac{2}{\mu_0 \gamma f Q M_0}.$$

Substitute $\mu_0 = 4\pi*10^{-7}$ T m/A:

$$T_{rd} = \frac{1}{2\pi*10^{-7} \gamma f Q M_0}.$$

Convert $\gamma$ from Hz/T to MHz/T with the conversion factor 1 MHz = $10^6$ Hz.

$$T_{rd} = \frac{1}{2\pi*10^{-7}(\gamma*10^6) f Q M_0}, \text{ which becomes}$$

$$T_{rd} = \frac{1}{2\pi*10^{-1} \gamma f Q M_0}.$$

Notice the extra factor of $10^{-1}$ in the denominator of Eq. 4 as compared with Eq. 1. Since these equations are now using the same units (MHz/T for $\gamma$, SI for all other units) they do not produce identical results.